\documentclass[a4paper,11pt]{article}
\usepackage{pos}
\usepackage{subfigure}

\title{Charmed baryon decays at BESIII}

\author*[a]{Cheng-Zhi He}
\onbehalf{On behalf of the BESIII Collaboration}

\affiliation[a]{School of Physics, Peking University,\\
  209 Chengfu Road, Beijing, People’s Republic of China}

\emailAdd{hechengzhi@pku.edu.cn}
\emailAdd{1900011335@pku.edu.cn}

\abstract{
BESIII has accumulated 6.4~fb$^{-1}$ of $e^+e^-$ collision data at center-of-mass energies from 4.60 to 4.95~GeV, corresponding to the world's largest sample of $\Lambda_c^+\bar\Lambda_c^-$ pairs. This proceeding summarizes recent BESIII results on charmed baryon decays, including the first observation of the transverse polarization of the $\Lambda_c^+$ in $e^+e^-\to\Lambda_c^+\bar\Lambda_c^-$, a precision measurement of the inclusive decay $\Lambda_c^+\to\Lambda X$, the first observations of $\Lambda_c^+\to ne^+\nu_e$ and $\Lambda_c^+\to n\pi^+\eta$ with deep-learning techniques, the first observation of $\Lambda_c^+\to pK^-e^+\nu_e$ together with evidence for $\Lambda_c^+\to\Sigma^\pm\pi^\mp e^+\nu_e$, and partial wave analyses of $\Lambda_c^+\to\Lambda\pi^+\eta$.
}

\FullConference{43rd International Conference on High Energy Physics (ICHEP 2026)\\
30 July  to 5 August , 2026\\
Natal, Brazil\\}

\begin{document}
\maketitle

\section{Introduction}
\label{sec:intro}

The $\Lambda_c^+$ is the lowest-lying charmed baryon, consisting of a heavy charm quark and an unexcited, zero-spin $ud$ diquark. Since most charmed baryons eventually decay into the $\Lambda_c^+$, it plays a central role in heavy-flavour spectroscopy and in the study of charmed baryon dynamics.

The theoretical description of charmed baryon decays remains challenging, since the relevant energy scale lies between the perturbative and non-perturbative regimes of QCD. In contrast to charmed meson decays, the non-factorizable $W$-exchange amplitudes in $\Lambda_c^+$ decays are neither colour suppressed nor helicity suppressed, and can be comparable to or even larger than the factorizable contributions. A number of phenomenological approaches have therefore been developed, including the factorization scheme, $SU(3)$ flavour symmetry, topological diagrams and so on. Measurements of $\Lambda_c^+$ decays provide essential tests of these approaches and important inputs to theory.

The BESIII detector~\cite{besiii_detector} records symmetric $e^+e^-$ collisions delivered by the BEPCII storage ring at center-of-mass energies from 1.84 to 4.95~GeV with a peak luminosity of $1.1\times10^{33}$~cm$^{-2}$s$^{-1}$. So far, BESIII has collected about 6.4~fb$^{-1}$ of $e^+e^-$ annihilation data at 13 energy points from $\sqrt{s}=4.60$ to 4.95~GeV \cite{besiii_charmdata,besiii_lineshape}, corresponding to about one million $\Lambda_c^+\bar\Lambda_c^-$ pairs. This is the world's largest $\Lambda_c^+\bar\Lambda_c^-$ sample and enables precision studies of $\Lambda_c^+$ production and decays.


\section{Transverse polarization and decay dynamics of the \texorpdfstring{$\Lambda_c^+$}{Lambda(c)+}}
\label{sec:polarization}

In $e^+e^-$ annihilation near the $\Lambda_c^+\bar\Lambda_c^-$ threshold, the pair production is dominated by single-photon exchange, which is described by two complex electromagnetic form factors (EMFFs), $G_E$ and $G_M$. A non-zero relative phase $\Delta\Phi$ between them gives rise to a transverse polarization of the $\Lambda_c^+$ and provides additional constraints on the decay dynamics.

\begin{figure}[htbp]
\centering
\subfigure[]{\includegraphics[width=0.4\textwidth]{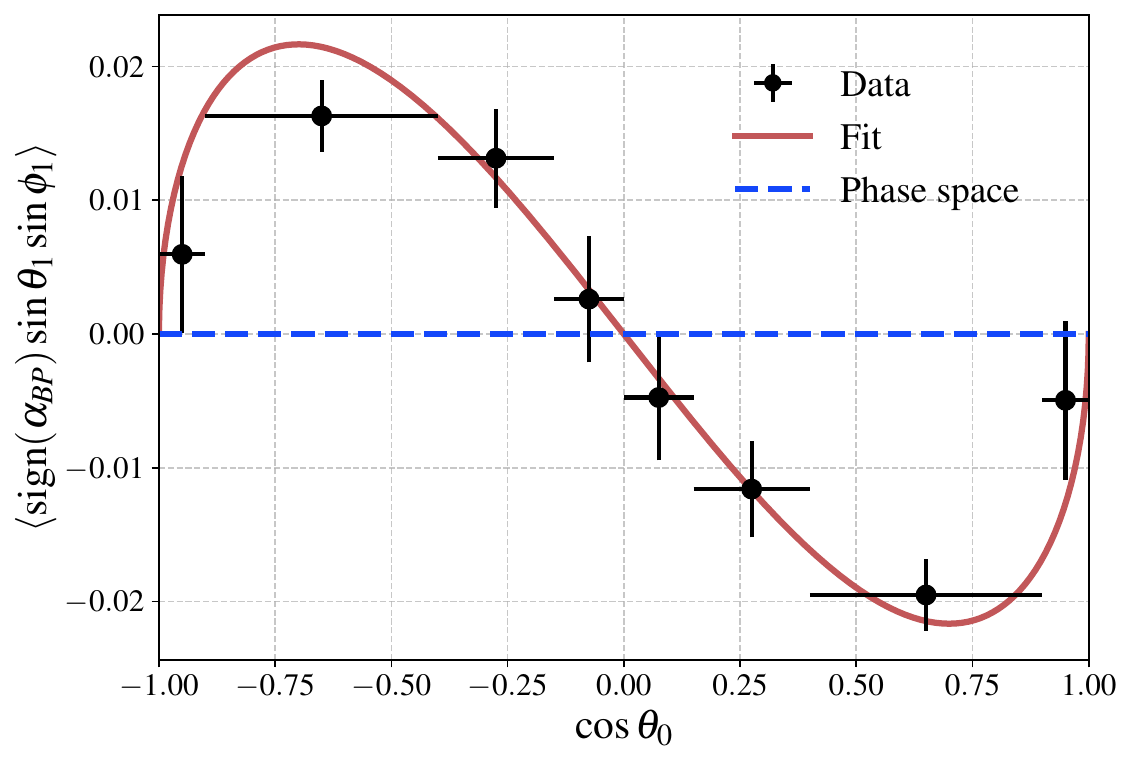}}
\subfigure[]{\includegraphics[width=0.4\textwidth]{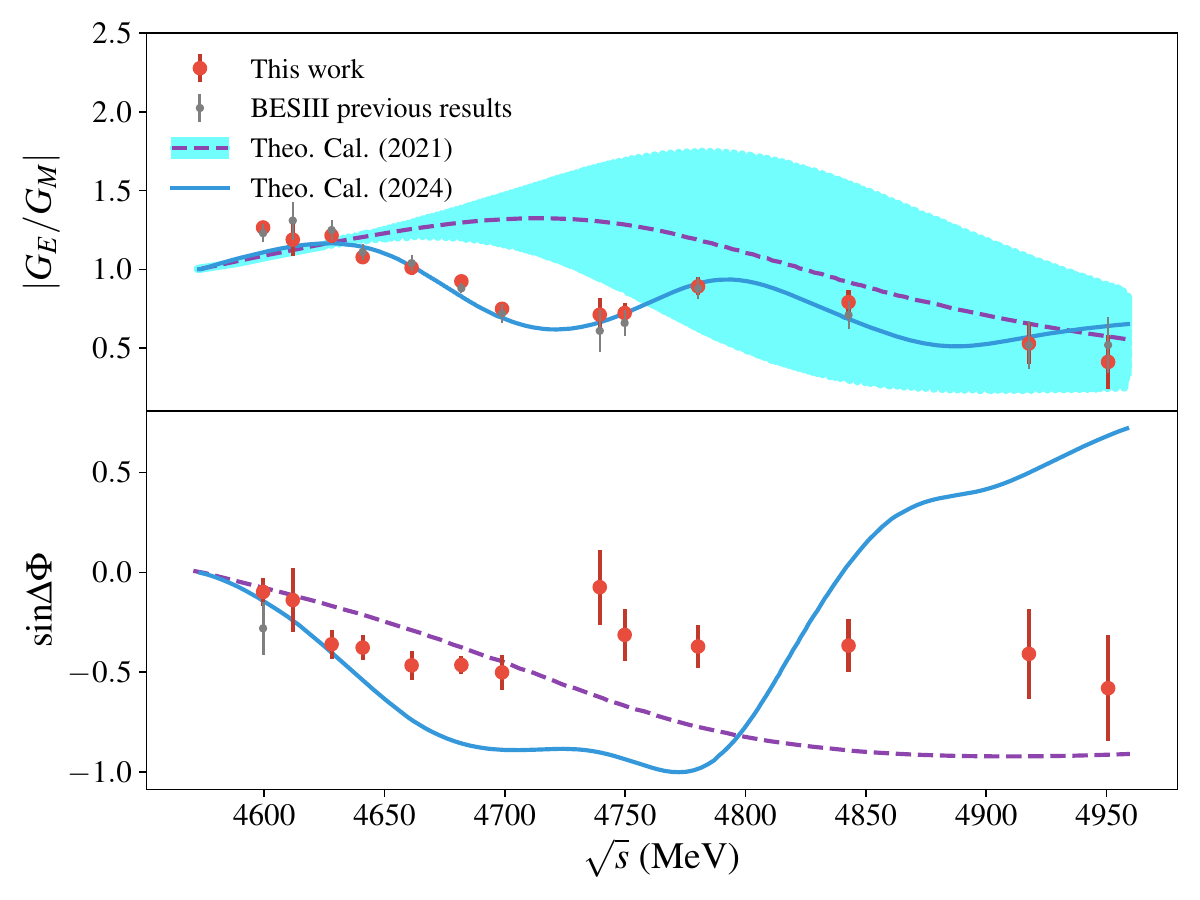}}
\caption{(a) Polarization moment as a function of $\cos\theta$. The red curve is the global fit including transverse polarization and the blue dashed line is the expectation without polarization. (b) Comparison of the measured ratio $|G_E/G_M|$ (upper panel) and $\sin\Delta\Phi$ (lower panel) with theoretical calculations~\cite{theo1,theo2}.}
\label{fig:pol}
\end{figure}

Five decay modes are used as signal channels: $\Lambda_c^+\to\Lambda\pi^+$, $\Lambda_c^+\to pK_S^0$, $\Lambda_c^+\to\Sigma^0\pi^+$, $\Lambda_c^+\to\Sigma^+\pi^0$ and $\Lambda_c^+\to pK^-\pi^+$. A simultaneous analysis of the joint angular distributions of these modes leads to the first observation of the transverse polarization of the $\Lambda_c^+$ in $e^+e^-\to\Lambda_c^+\bar\Lambda_c^-$~\cite{besiii_polarization}, as shown in Fig.~\ref{fig:pol}. The measured $\Delta\Phi$ deviates from the theoretical prediction, indicating that a better understanding of the electromagnetic form factors is still required.

The same data set allows the decay asymmetry parameters of the four two-body decay modes to be determined, together with the $S$- and $P$-wave phase shift $\delta_p-\delta_s$ and the $CP$ asymmetry
\begin{equation}
  A_{CP}^{\alpha_{BP}} = \frac{\alpha_{BP}+\bar{\alpha}_{BP}}{\alpha_{BP}-\bar{\alpha}_{BP}} ,
\end{equation}
where $\alpha_{BP}$ ($\bar{\alpha}_{BP}$) is the decay asymmetry parameter of the $\Lambda_c^+$ ($\bar\Lambda_c^-$) decay. The results are summarized in Table~\ref{tab:asym}~\cite{besiii_polarization}.

\begin{table}[htbp]
  \centering
  \caption{Decay asymmetry parameters $\alpha_{BP}$, the $S$--$P$ wave phase shift
  $\delta_p-\delta_s$ and the $CP$ asymmetry $A_{CP}^{\alpha_{BP}}$ measured in the four
  Cabibbo-favoured decays. The first uncertainty is statistical and the second is
  systematic.}
  \label{tab:asym}
  \begin{tabular}{l|ccc}
  \hline\hline
  Decay & $\alpha_{BP}$ & $\delta_p-\delta_s$ & $A_{CP}^{\alpha_{BP}}$ \\
  \hline
  $\Lambda_c^+\to pK_S^0$ & $-0.918^{+0.133}_{-0.082}\pm0.031$ & / & $0.079^{+0.115}_{-0.101}\pm0.019$ \\
  $\Lambda_c^+\to\Lambda\pi^+$ & $-0.790\pm0.032\pm0.009$ & $2.71^{+0.28}_{-0.17}\pm0.02$ & $0.002\pm0.047\pm0.017$ \\
  $\Lambda_c^+\to\Sigma^0\pi^+$ & $-0.502\pm0.080\pm0.009$ & $2.19^{+0.49}_{-0.13}\pm0.02$ & $0.206^{+0.188}_{-0.156}\pm0.028$ \\
  $\Lambda_c^+\to\Sigma^+\pi^0$ & $-0.590\pm0.049\pm0.022$ & $2.23^{+0.19}_{-0.06}\pm0.02$ & $-0.086\pm0.081\pm0.029$ \\
  \hline\hline
  \end{tabular}
\end{table}

No evidence for $CP$ violation is found in the four CF decays. The phase shifts $\delta_p-\delta_s$ are large, indicating that final-state rescattering effects are important and may enhance $CP$ violation in Cabibbo-suppressed decays.

\section{Precision study of the inclusive decay \texorpdfstring{$\Lambda_c^+\to\Lambda X$}{Lc -> Lambda X}}
\label{sec:lambdax}

The inclusive decay $\Lambda_c^+\to\Lambda X$ is studied with the double-tag technique, in which the $\bar\Lambda_c^-$ is reconstructed on the tag side and a $\Lambda$ together with the remaining particles on the signal side. The absolute branching fraction is measured to be $(38.07\pm0.38\pm0.49)\%$, consistent with the PDG value of $(38.2^{+2.9}_{-2.4})\%$~\cite{pdg} and improving the precision by about a factor of four~\cite{besiii_lambdax}. Combined with the known exclusive decays, this leaves $(4.48\pm1.25)\%$ of the exclusive decays unmeasured.

In addition, the longitudinal polarization of the $\Lambda$ is measured for the first time in an inclusive $\Lambda_c^+$ decay, giving $\mathcal{P}=-0.393\pm0.055\pm0.020$ for the $\Lambda$ and $\bar{\mathcal{P}}=0.288\pm0.056\pm0.017$ for the $\bar\Lambda$. Two $CP$ observables are constructed: the direct asymmetry $A_{CP}^{\rm dir}=(1.5\pm1.0\pm1.0)\%$ and the polarization asymmetry
\begin{equation}
  A_{CP}^{\rm pol} \equiv
  \frac{\mathcal{P}_\Lambda\alpha_- - \mathcal{P}_{\bar\Lambda}\alpha_+}
       {\mathcal{P}_\Lambda\alpha_- + \mathcal{P}_{\bar\Lambda}\alpha_+}
  = 0.15\pm0.12\pm0.03 .
\end{equation}
No evidence for $CP$ violation is found in either of them.

\section{Semi-leptonic decays}
\label{sec:semi}

\subsection{First observation of \texorpdfstring{$\Lambda_c^+\to ne^+\nu_e$}{Lc -> n e nu}}

The semi-leptonic decay $\Lambda_c^+\to ne^+\nu_e$ provides access to the Cabibbo-Kobayashi-Maskawa matrix element $|V_{cd}|$ in the baryon sector. Since the signal process with a prompt neutron is difficult to distinguish from $\Lambda_c^+\to\Lambda e^+\nu_e(\Lambda\to n\pi^0)$ background by conventional methods. A deep-learning technique is therefore developed \cite{besiii_nenu}: a graph neural network (GNN) is used to classify the energy-deposition patterns of neutrons and $\Lambda$ baryons in the electromagnetic calorimeter, together with a data-driven pipeline for the GNN training, calibration and validation and for the quantification of the systematic uncertainties.

\begin{figure}[htbp]
\centering
\subfigure[]{\includegraphics[width=0.6\textwidth]{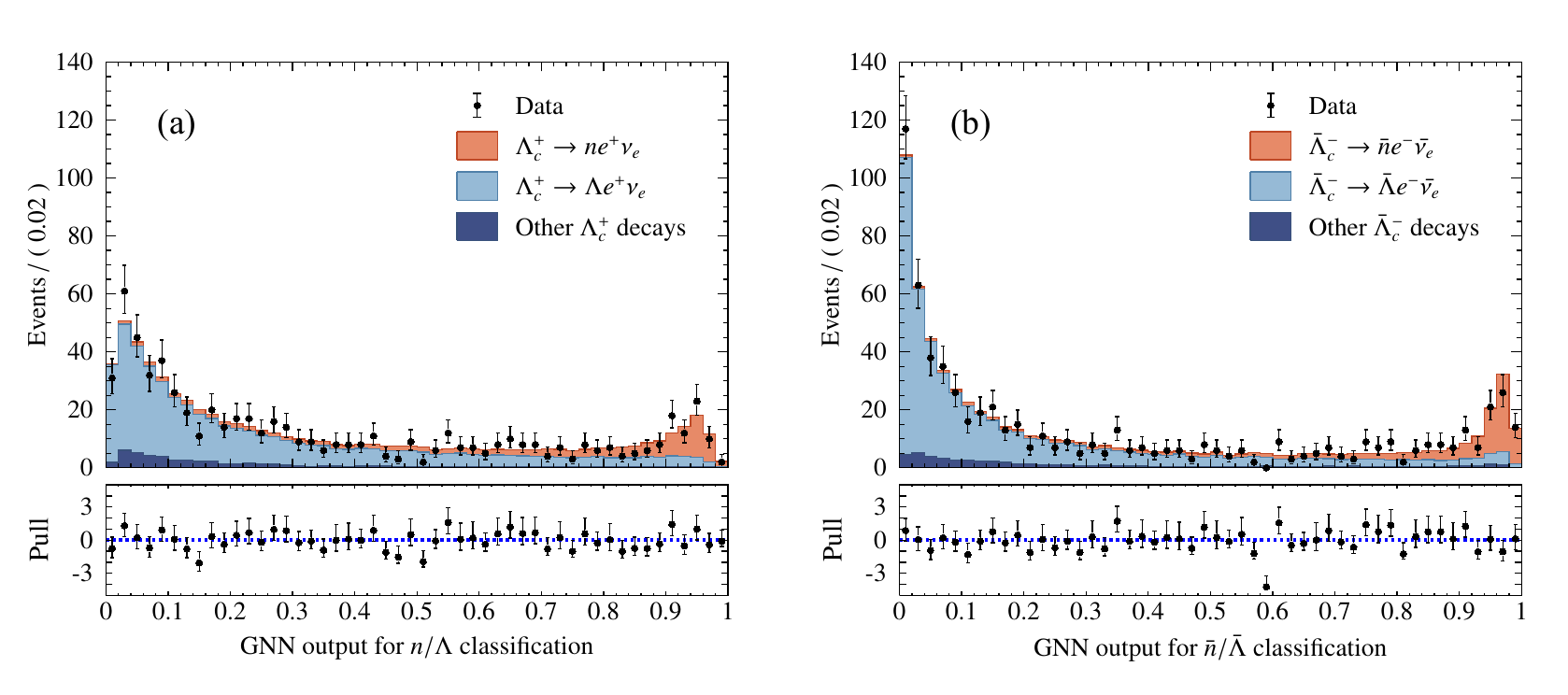}}
\subfigure[]{\includegraphics[width=0.35\textwidth]{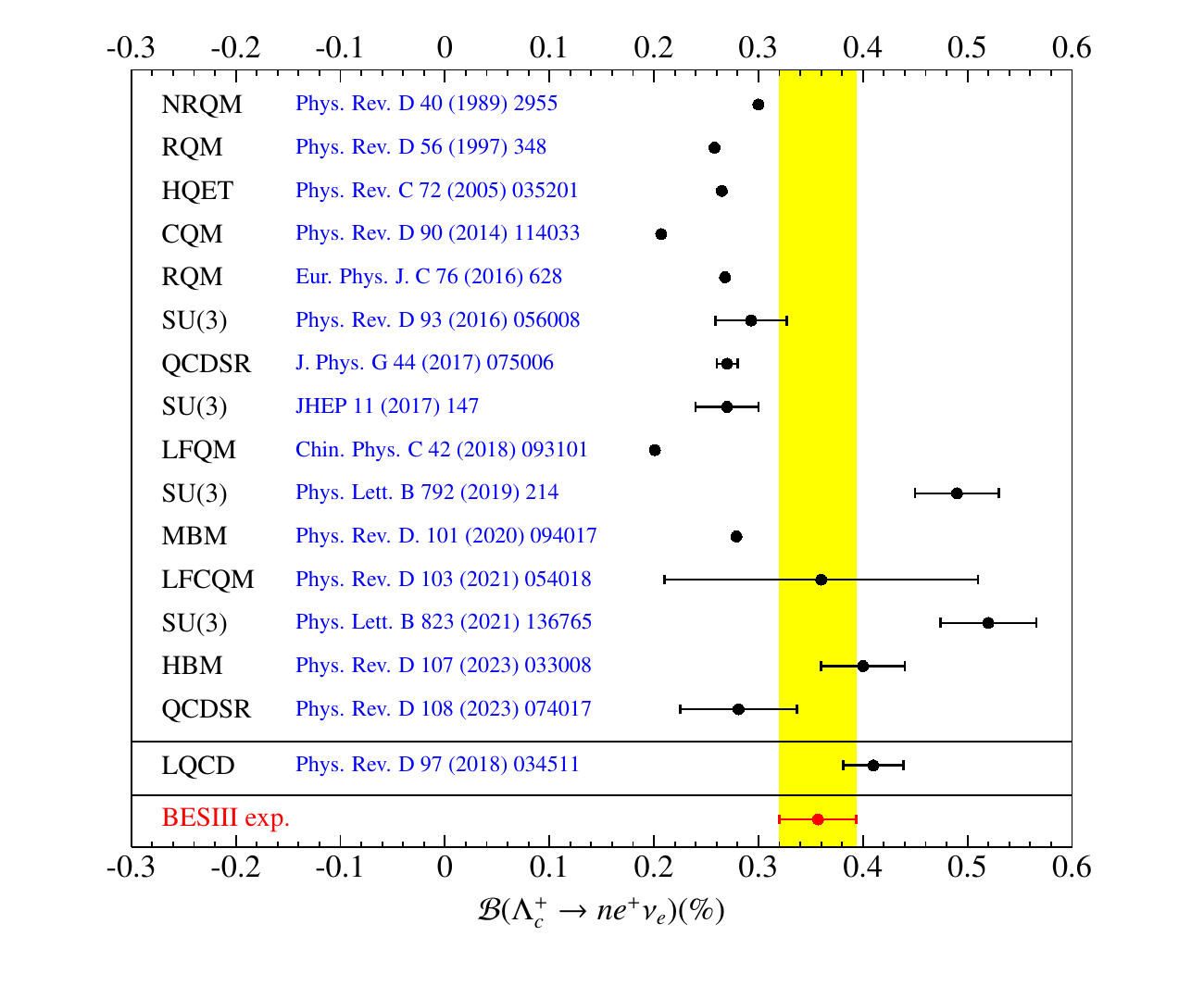}}
\caption{ (a) GNN-output distributions for $\Lambda_c^+\to n(\Lambda)e^+\nu_e$ and the charge-conjugate processes in data. (b) Comparison of the BESIII result with theoretical predictions.}
\label{fig:gnn}
\end{figure}

As shown in Fig.~\ref{fig:gnn}, the corrected GNN-score distribution is used to extract the signal yield. The decay $\Lambda_c^+\to ne^+\nu_e$ is observed for the first time with a statistical significance of more than $10\sigma$, and its absolute branching fraction is measured to be $(0.357\pm0.033_{\rm stat.}\pm0.014_{\rm syst.})\%$. Using lattice-QCD form factors as input, this leads to the first determination of $|V_{cd}|$ from charmed-baryon decays, $|V_{cd}| = 0.208\pm0.011_{\rm exp.}\pm0.007_{\rm LQCD}\pm0.011_{\tau_{\Lambda_c^+}}$.

\subsection{\texorpdfstring{$\Lambda_c^+\to pK^-e^+\nu_e$}{Lc -> pK e nu} and
\texorpdfstring{$\Lambda_c^+\to\Sigma^\pm\pi^\mp e^+\nu_e$}{Lc -> Sigma pi e nu}}

The semi-leptonic decay $\Lambda_c^+\to pK^-e^+\nu_e$ is observed for the first time with
a significance of $8.2\sigma$ \cite{besiii_pkepnue}, and its branching fraction is
measured to be $(0.88\pm0.17_{\rm stat.}\pm0.07_{\rm syst.})\times10^{-3}$. Evidence is also found for the intermediate processes
$\Lambda_c^+\to\Lambda(1405)e^+\nu_e$ ($3.2\sigma$) and
$\Lambda_c^+\to\Lambda(1520)e^+\nu_e$ ($3.3\sigma$).

More recently, evidence for $\Lambda_c^+\to\Sigma^\pm\pi^\mp e^+\nu_e$ has been reported with a significance of $3.6\sigma$ \cite{besiii_sigpienu}, with a branching fraction of $(7.7^{+2.5}_{-2.3\,\rm stat.}\pm1.3_{\rm syst.})\times10^{-4}$. A future joint analysis of the two channels, $\Lambda_c^+ \to p K^- e^- \nu_e$ and $\Sigma \pi^+ e^- \nu_e$, can help discover the $\Lambda(1405)/(1520)e^+\nu_e$ processes, and understand the dynamical properties of $\Lambda_c^+\to\Lambda^*e^+\nu_e$.

\section{First observation of \texorpdfstring{$\Lambda_c^+\to n\pi^+\eta$}{Lc -> n pi eta}}
\label{subsec:npieta}

The decay $\Lambda_c^+\to n\pi^+\eta$ is observed for the first time \cite{besiii_npieta}. Because of the large hadronic backgrounds, a deep-learning approach based on a Transformer architecture is employed to separate the signal from the background, and the observation has a statistical significance of $9.5\sigma$. The ratio of branching fractions is measured to be
\begin{equation}
  \frac{\mathcal{B}(\Lambda_c^+\to n\pi^+\eta)}{\mathcal{B}(\Lambda_c^+\to\Lambda\pi^+\eta)}
  = 0.155\pm0.031_{\rm stat.}\pm0.012_{\rm syst.} .
\end{equation}
Taking the world average of $\mathcal{B}(\Lambda_c^+\to\Lambda\pi^+\eta)$ as reference, the absolute branching fraction is determined to be $(2.94\pm0.59_{\rm stat.}\pm0.23_{\rm syst.}\pm0.13_{\rm ref.})\times10^{-3}$. The intermediate process $\Lambda_c^+\to n a_0(980)^+$ is also searched for in the $\pi^+\eta$ invariant mass spectrum. No significant signal is found, and the upper limit at the $90\%$ confidence level is
\begin{equation}
  \mathcal{B}(\Lambda_c^+\to n a_0(980)^+)\times\mathcal{B}(a_0(980)^+\to\pi^+\eta)
  < 8.4\times10^{-4} .
\end{equation}

\section{Partial wave analysis of $\Lambda_c^+\to\Lambda\pi^+\eta$}
\label{sec:pwa}

\begin{figure}[htbp]
\centering
\includegraphics[width=0.32\textwidth]{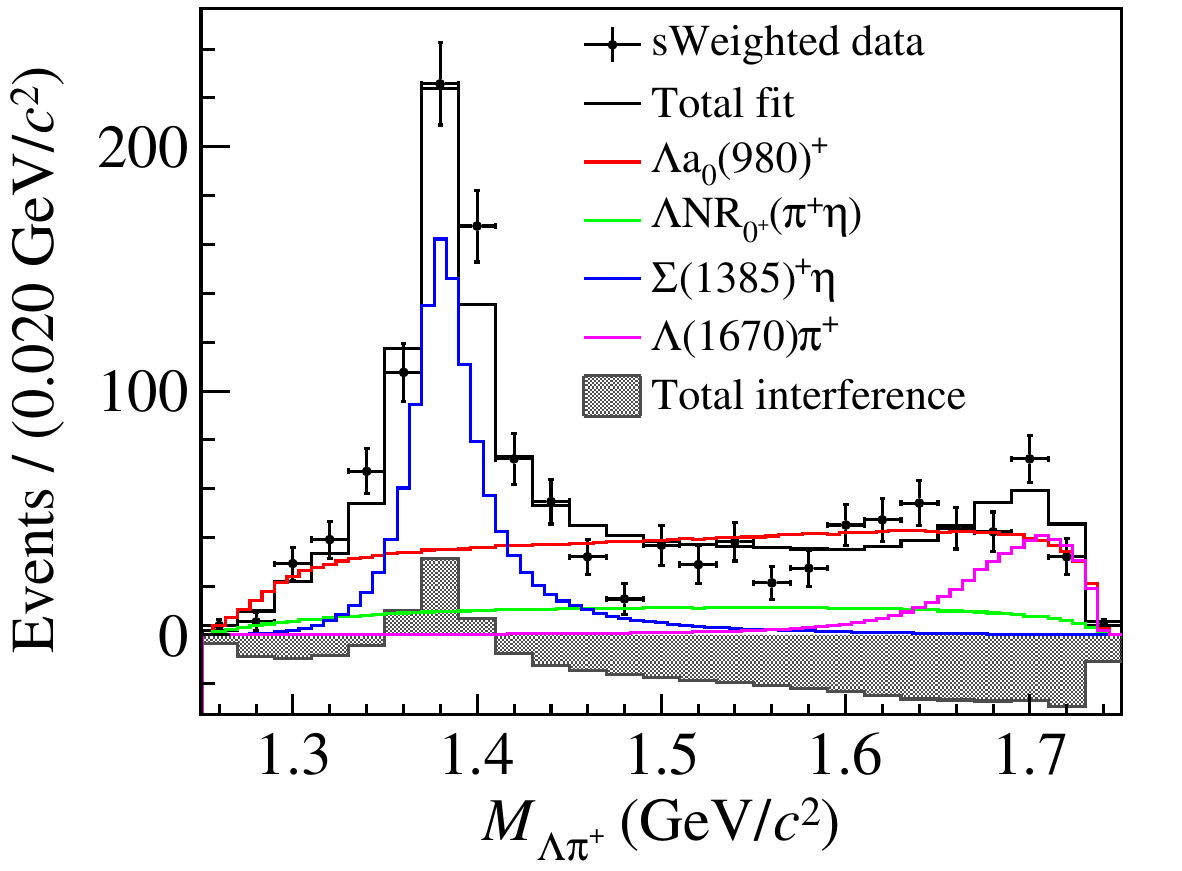}
\includegraphics[width=0.32\textwidth]{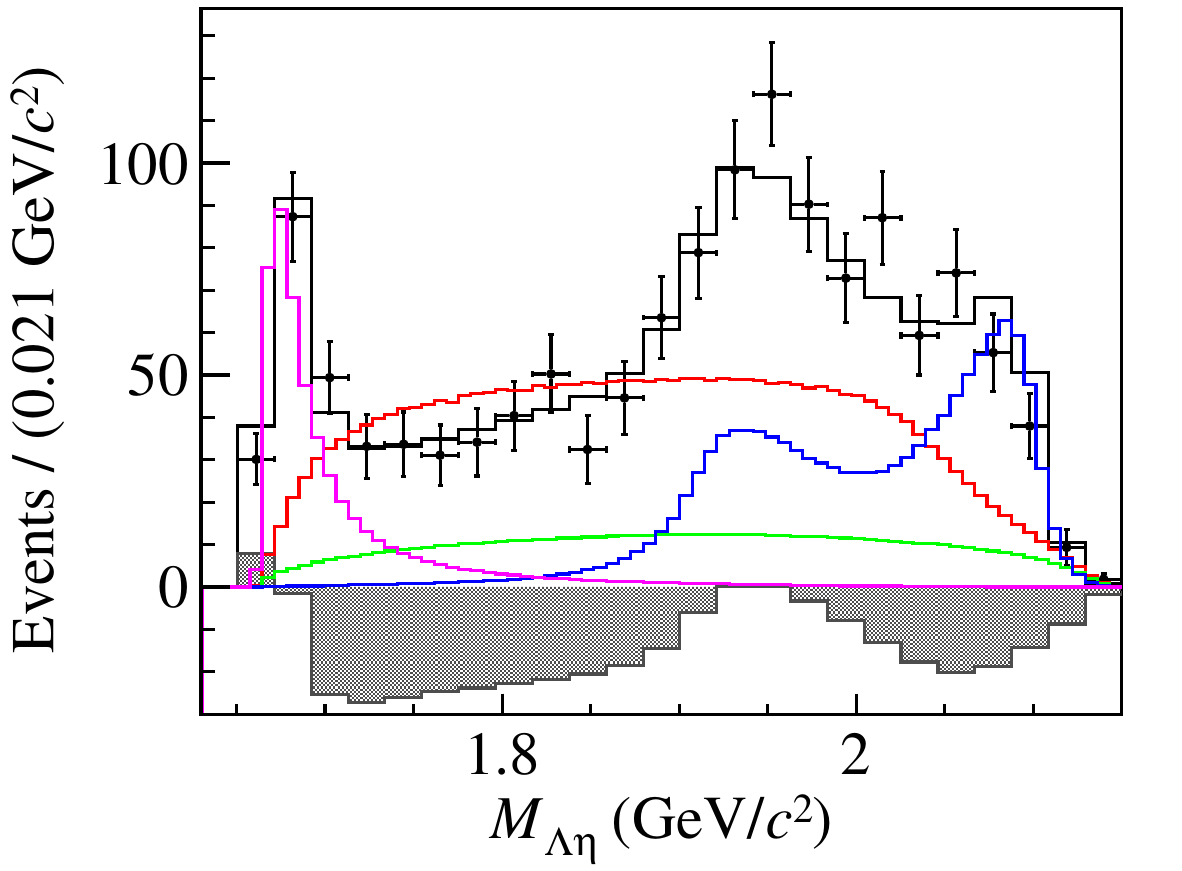}
\includegraphics[width=0.32\textwidth]{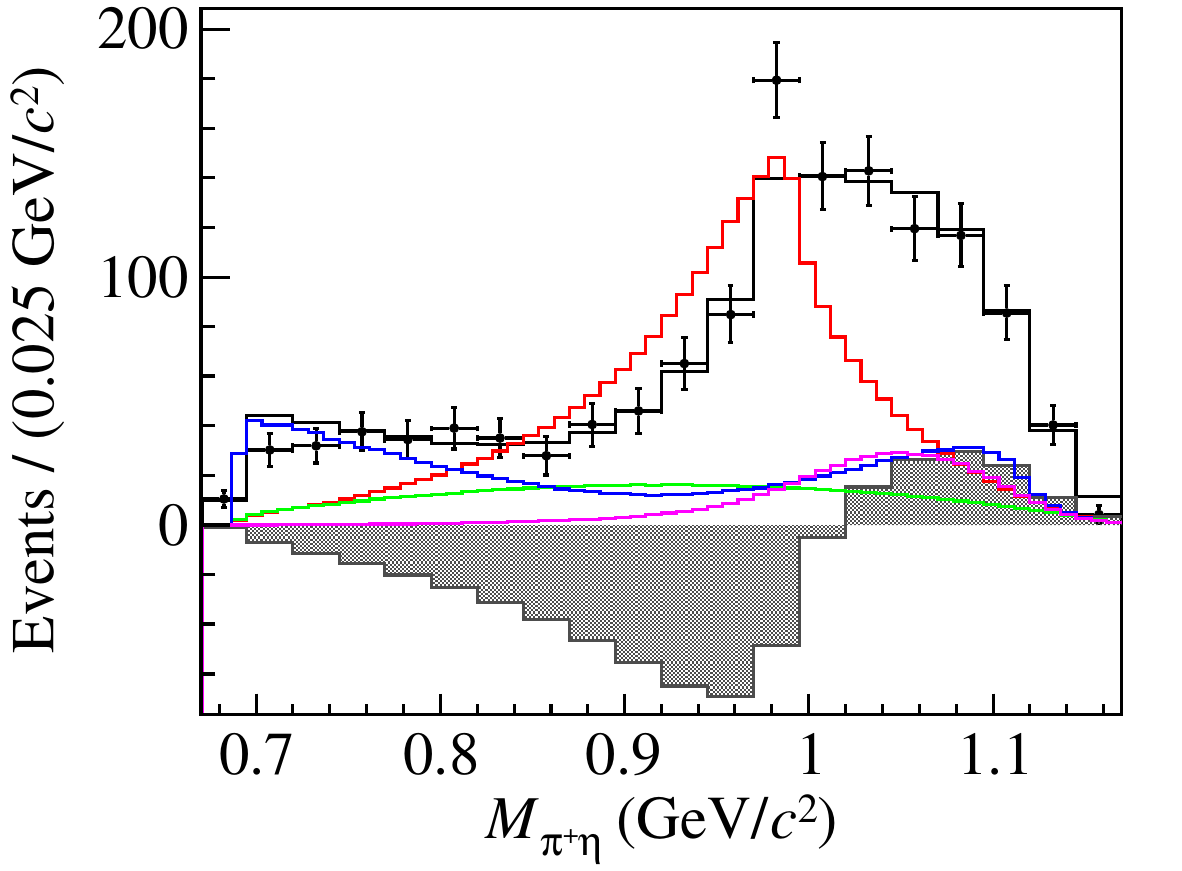}
\caption{projections of the fit results for $\Lambda_c^+\to\Lambda\pi^+\eta$ in the invariant mass spectra $M_{\Lambda\pi^+}$, $M_{\Lambda\eta}$ and $M_{\pi^+\eta}$}
\label{fig:pwa}
\end{figure}

A partial wave analysis (PWA) of the decay $\Lambda_c^+\to\Lambda\pi^+\eta$ is performed using the helicity amplitude formalism \cite{besiii_lmpieta}. The projections of fit results onto the invariant mass spectra are shown in Fig.~\ref{fig:pwa}. The absolute branching fraction is measured to be $(1.94\pm0.07\pm0.11)\%$. The intermediate process $\Lambda_c^+\to\Lambda a_0(980)^+$ is observed for the first time with a significance of $13.1\sigma$, and
\begin{equation}
  \mathcal{B}(\Lambda_c^+\to\Lambda a_0(980)^+)\times\mathcal{B}(a_0(980)^+\to\pi^+\eta)
  = (1.05\pm0.16_{\rm stat.}\pm0.05_{\rm syst.}\pm0.07_{\rm ext.})\% ,
\end{equation}
which exceeds the theoretical calculations by one to two orders of magnitude. Evidence for the potential pentaquark state $\Sigma(1380)^+$ is also found, with a significance of $6.1\sigma$ ($3.3\sigma$) when the $\pi^+\eta$ non-resonant contribution is excluded (included).

\section{Summary and outlook}
\label{sec:summary}

BESIII has collected $6.4$~fb$^{-1}$ of $e^+e^-$ annihilation data at center-of-mass energies between 4.60 and 4.95~GeV, providing the largest $\Lambda_c^+\bar\Lambda_c^-$ data set available to date. With these data, a broad set of $\Lambda_c^+$ decay properties has been investigated over the past two years. The transverse polarization of the $\Lambda_c^+$ in threshold production is observed for the first time and the associated decay parameters are extracted, while $CP$ violation is searched for in the four CF decays and in the inclusive decay $\Lambda_c^+\to\Lambda X$, where the $\Lambda$ longitudinal polarization is measured for the first time in an inclusive charmed-baryon decay. The decays $\Lambda_c^+\to ne^+\nu_e$, $\Lambda_c^+\to pK^-e^+\nu_e$, $\Lambda_c^+\to\Sigma^\pm\pi^\mp e^+\nu_e$ and $\Lambda_c^+\to n\pi^+\eta$ are observed or found as evidence, and a partial wave analysis of $\Lambda_c^+\to\Lambda\pi^+\eta$ involving potential light resonances is performed. BEPCII-U upgrade and new $\Lambda_c^+$ data taking is in progress, which will further extend these studies.

\end{document}